\documentclass[letterpaper,journal]{IEEEtran}

\usepackage[utf8]{inputenc}
\usepackage{amsmath, amsfonts, amsthm, amssymb}
\usepackage[ruled,linesnumbered]{algorithm2e}
\usepackage{graphicx}
\usepackage{tikz}
\usetikzlibrary{arrows.meta}
\usepackage{cite}
\usepackage{url}
\usepackage{textcomp}
\usepackage{xcolor}
\usepackage{soul}
\sethlcolor{yellow}
\soulregister\cite7
\soulregister\ref7
\soulregister\eqref7
\newcommand{\rev}[1]{#1}          
\usepackage{algorithmic}

\newcommand{\Heff}{\mathbf{H}_{\text{eff}}}

\newcommand{\Rsum}{R_{\text{sum}}}
\newcommand{\zobs}{z_{\text{obs}}}
\newcommand{\xobs}{x_{\text{obs}}}

\usepackage{stfloats} 
\usepackage{array}
\usepackage[caption=false,font=normalsize,labelfont=sf,textfont=sf]{subfig}
\usepackage{verbatim}
\def\BibTeX{{\rm B\kern-.05em{\sc i\kern-.025em b}\kern-.08em
    T\kern-.1667em\lower.7ex\hbox{E}\kern-.125emX}}
\usepackage{balance}
\usepackage{bm}
\usepackage{threeparttable}

\begin{document}

\title{Overcoming Half-Space Blockage: Airy Beamforming for Radiative Near-Field Multiuser XL-MIMO Communications}

\author{Yifeng Qin,~\IEEEmembership{Member,~IEEE,}
        Jing Chen,
        Zhi Hao Jiang,~\IEEEmembership{Member,~IEEE,}
        Zhi Ning Chen,~\IEEEmembership{Fellow,~IEEE,}\\
        Yongming Huang,~\IEEEmembership{Fellow,~IEEE,}
        and~Lingyang Song,~\IEEEmembership{Fellow,~IEEE}
\thanks{This work was supported by ... (Corresponding author: Yifeng Qin)}
\thanks{Y. Qin and J. Chen are with the Peng Cheng Laboratory, Shenzhen, 518052, China (e-mails: ee06b147@gmail.com, chenj12@pcl.ac.cn).}
\thanks{Z. H. Jiang is with the State Key Laboratory of Millimeter Waves, School of Information Science and Engineering, Southeast University, Nanjing 210096, China (e-mail: zhihao.jiang@seu.edu.cn).}
\thanks{Zhi Ning Chen is with the Department of Electrical and Computer Engineering, National University of Singapore, Singapore, 117583 (e-mail: eleczn@nus.edu.sg).}
\thanks{Yongming Huang is with the National Mobile Communication Research Laboratory, and the School of Information Science and Engineering, Southeast University, Nanjing 210096, China, and also with the Purple Mountain Laboratories, Nanjing 211111, China (e-mail: huangym@seu.edu.cn).}
\thanks{L. Song is with the Peng Cheng Laboratory, Shenzhen, 518052, and School of Electrical and Computer Science, Peking University, Beijing, 100871, China (e-mail: lingyang.song@pku.edu.cn; songly@pcl.ac.cn).}
}

\markboth{IEEE Transactions on Wireless Communications,~Vol.~XX, No.~X, October~2025}%
{Qin \MakeLowercase{\textit{et al.}}: Overcoming Half-Space Blockage: Airy Beamforming for RNF Multiuser XL-MIMO}


\maketitle

\begin{abstract}
\rev{Next-generation wireless systems with extremely large-scale antenna arrays (ELAAs) operate in the radiative near-field (RNF) region, where distance-aware focusing is feasible but is easily interrupted by half-space obstacles such as walls and corners. When the aperture-to-user path is clipped by a knife edge, conventional focused beams are truncated and the multi-user channel suffers a collapse of its weak singular mode that defeats zero-forcing (ZF) precoding. In this paper, a hybrid transmission strategy is proposed in which a low-dimensional Airy radio-frequency beam family, parameterized by only three physical quantities (launch angle, virtual focal length, and signed curvature), is used to carry energy around an accessible diffracting edge, while a digital ZF stage performs the final inter-user cancellation under one explicitly declared total-radiated-power constraint. Channel separability is evaluated on the whitened effective channel, so that non-orthogonal analog beams cannot distort the comparison. In the principal blocked two-user scene, the proposed design attains 4.94 bit/s/Hz, compared with 1.97 for conventional focusing, 4.61 for a caustic construction, 5.93 for unrestricted digital ZF, and 6.47 for an iterative fully digital WMMSE reference, and the whitened smallest singular value is more than doubled. Across 36 blocked-user geometries, the whitened conditioning and the minimum user rate are improved in every case. The approach requires an accessible diffracting edge and does not restore service through a fully opaque plane.}
\end{abstract}


\section{Introduction}
\label{sec:introduction}

\IEEEPARstart{T}{HE} advent of 6G wireless systems is characterized by the deployment of Extremely Large-Scale Antenna Arrays (ELAA) operating at millimeter-wave (mmWave) and Terahertz (THz) frequencies \cite{ref1, ref2, ref3}. These massive apertures extend the Radiative Near-Field (RNF) to hundreds of meters, unlocking the ``distance dimension'' for precise beam focusing \cite{ref4, ref5}.

\begin{figure*}[!t]
    \centering
    \includegraphics[width=\textwidth]{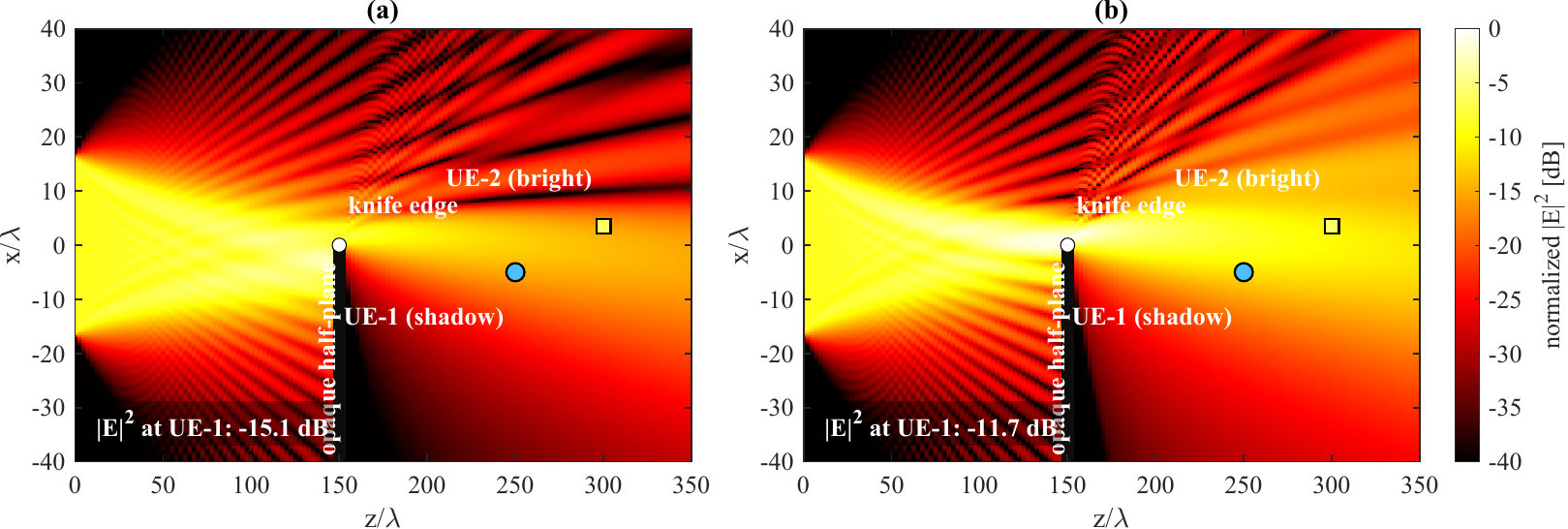}
    \caption{\rev{Physical scope and design principle, rendered from fields computed with the paper's angular-spectrum model. (a)~A conventional beam focused on UE-1 is clipped by the opaque half-plane at $z=150\lambda$ and delivers $-15.1$~dB of normalized intensity at the shadowed user. (b)~The optimized Airy beam of Sec.~V, $(B,F,\Delta\theta)=(+10,\,1.9~\mathrm{m},\,+1.9^{\circ})$, redistributes aperture energy through the open half-plane and delivers $-11.7$~dB at the same point while continuing to serve UE-2. The bright curved feature is an intensity ridge of the diffracted field: energy reaches the shadow by diffraction around the accessible knife edge, and a fully opaque plane with no open path leaves every transmitter-side scheme in outage (Sec.~IV-D).}}
    \label{fig:Conceptual}
\end{figure*}

However, high-frequency RNF links are intrinsically fragile. As visually summarized in Fig.~\ref{fig:Conceptual}, ubiquitous half-space obstacles (e.g., walls) create extensive geometric shadow regions. In these scenarios, traditional near-field beams, which propagate linearly \cite{ref6, ref7, ref8}, are severely truncated by the obstacle edge. This results in a catastrophic power loss and a ``singular value collapse'' of the multi-user channel \cite{ref9}, often rendering Zero-Forcing precoding ineffective and causing service outage.

To mitigate such blockage, Reconfigurable Intelligent Surfaces (RIS) are often employed to establish virtual Line-of-Sight (LoS) links \cite{ref11, ref12, ref13, ref14}. Yet, ubiquitous RIS deployment incurs significant hardware costs, power consumption, and channel estimation overhead \cite{ref7}. This motivates a complementary line of work on \textit{Wavefront Engineering}---shaping the electromagnetic field structure at the transmitter to navigate complex environments \rev{when no auxiliary reflecting aperture is available. We emphasize at the outset that the two approaches address different path mechanisms and are complementary rather than interchangeable: transmitter-side shaping exploits an accessible diffracting edge of the obstruction itself, whereas RIS assistance creates a separate reflected route (Sec.~V-G).}

\rev{The progression from far-field beamforming to near-field beamfocusing sharpens what an aperture can control, but it does not remove the dependence on an unobstructed path. Far-field beam steering controls only the angular direction of a quasi-planar wavefront; RNF beamfocusing additionally exploits wavefront curvature to resolve users in range \cite{ref9,SinghSPM}. Both, however, transport energy along essentially straight ray bundles, so a knife edge that clips the aperture-to-user path degrades them in the same way. This limitation motivates structured wavefronts whose intensity maxima follow curved trajectories \cite{SinghSPM}, of which the Airy family considered here is the canonical accelerating example.}

Recent years have witnessed a surge of interest in Non-Diffracting (ND) beams for this purpose. The seminal work by Reddy and Jornet \emph{et al.} \cite{Jornet2023Bessel} demonstrated the potential of THz Bessel beams to ``self-heal'' after encountering partial blockage, triggering a wave of research into structured light for wireless communications. Subsequent studies have expanded this frontier: Miao \emph{et al.} \cite{Miao2023Bessel} realized extended non-diffractive ranges using dielectric metalenses, while Yang \emph{et al.} \cite{Yang2024Flexible} explored flexible antenna arrays to dynamically shape radiation patterns. In our recent work \cite{Qin2025Roadmap}, we provided a quantitative roadmap for the engineering of linear ND beams (specifically Bessel and Mathieu beams) using planar phased arrays, establishing the fundamental link-level gain regimes for these structured wavefronts. Complementing these theoretical studies, Bodet \emph{et al.} \cite{Bodet2024Measurements} provided essential sub-terahertz near-field channel measurements, verifying the propagation characteristics of diffractive beams.

However, a critical limitation remains: Bessel and Mathieu beams, while robust to partial occlusion, still propagate along \textit{linear} trajectories. They cannot reach deep into the geometric shadow of large, optically opaque obstacles. To address this, curving beams based on the Airy function have emerged as a promising solution. Guerboukha \emph{et al.} \cite{Guerboukha2024Curving} demonstrated the first experimental realization of curving THz wireless data links around obstacles. Subsequently, the potential of Airy beams has been rapidly explored across various domains: Darsena \emph{et al.} \cite{Darsena2025Fundamentals} analyzed their fundamental performance limits in near-field channels; Lee \emph{et al.} \cite{Lee2025Experimental} experimentally validated high-speed transmission using 3D-printed lenses; and Liu \emph{et al.} \cite{LiuConvex} proposed convex optimization frameworks for bending beam generation. In specific application scenarios, Zhao \emph{et al.} \cite{Zhao2025DataCenter} evaluated the efficacy of Airy beams for Terahertz wireless data centers, while Hu \emph{et al.} \cite{Hu2024NC} demonstrated robust underwater optical communications using RGB Airy transmitters. Beyond standard communications, the unique wavefronts of Airy beams have also been exploited for physical layer security via wavefront hopping \cite{Petrov2025Security}, complex spatiotemporal wavepacket dynamics \cite{Huang2025AiryAiry}, and high-resolution phase imaging through scattering media \cite{Wang2025PhaseImaging,Raza2025Reflective}.

Despite this flurry of activity, a critical gap persists in the literature. Most existing studies focus on single-user link budget enhancement \cite{Guerboukha2024Curving, Lee2025Experimental} or specific non-communication applications \cite{Petrov2025Security, Wang2025PhaseImaging}. The dynamics of \textit{Multi-User (MU)} interference when using these curved, diffractive beams remain largely unexplored. Unlike the linear Bessel beams discussed in \cite{Qin2025Roadmap}, Airy beams possess extended asymmetric sidelobes (the ``tails'' essential for self-bending), which pose a severe risk of co-channel interference in MU-MIMO setups. Standard linear precoding cannot easily resolve this conflict without sacrificing the bending gain.

In this paper, we propose a physics-aware Airy beamforming framework to bridge this gap. Our main contributions are:

\begin{itemize}
    \item \rev{\textbf{RNF Blockage Modeling under a Common Evaluation Contract:} We derive a Green's-function/angular-spectrum model of half-space blockage that reveals the singular-value collapse behind knife-edge obstacles, and we evaluate every scheme under one declared total-radiated-power constraint, with channel separability measured on the whitened effective channel so that non-orthogonal analog beams cannot distort the comparison.}
    \item \rev{\textbf{Edge-Riding Airy Beamforming with Quantified Breadth:} We design a three-parameter Airy analog family (launch angle, virtual focal length, signed curvature) with geometry-based initialization and local refinement. Under partial edge blockage it restores multi-user service where focusing fails; a stratified 77-scene blocker campaign quantifies where it helps (26 scenes), where it is equivalent (48), and where physics forbids service (3 outages), with controls and hard limits reported separately.}
    \item \rev{\textbf{Multiuser-Aware RF Shaping with Digital Cancellation:} We show that the optimized RF stage improves the whitened weak singular mode and the ZF power cost (in the principal scene, $\sigma_{\min}$ rises $2.24\times$ and the ZF cost falls by $78\%$), while the digital ZF stage performs the final interference cancellation; the RF stage acts as a conditioning stage rather than an analog null former. The design is benchmarked against constructive baselines re-implemented from the source papers and against fully digital references, reaching $76.3\%$ of an iterative WMMSE upper reference while giving the shadowed user a slightly higher rate.}
\end{itemize}

The remainder of this paper is organized as follows. Section~II establishes the RNF multi-user system model based on Green's function and introduces the Zero-Forcing precoding framework. Section~III analyzes the baseline performance of traditional near-field focusing in blockage-free scenarios, highlighting the distance-domain resolution of RNF. Section~IV investigates the critical impact of half-space blockage on channel singular values and demonstrates the rank restoration capability of the proposed edge-riding Airy beamforming. Section~V further explores a challenging mixed blockage scenario\rev{, presenting the three-parameter analog optimization together with fully digital and constant-modulus optimization references, per-user field evidence, and operating-boundary diagnostics}. Finally, Section~VI concludes the paper and discusses future research directions.

\section{System Model and Channel Characterization}
\label{sec:system_model}

We consider a downlink radiative near-field (RNF) multi-user MIMO (MU-MIMO) system operating in the millimeter-wave (mmWave) band. The Base Station (BS) employs a large-scale Uniform Linear Array (ULA) to serve $K$ single-antenna users simultaneously. The propagation environment is characterized by a half-space blockage geometry, where a knife-edge obstacle potentially obstructs the Line-of-Sight (LoS) paths for specific users.

\subsection{Geometry and Signal Configuration}
We define the system in a two-dimensional (2D) Cartesian coordinate system $(x, z)$, where the $z$-axis denotes the longitudinal propagation direction and the $x$-axis represents the transverse dimension.
The BS is equipped with an $N$-element ULA located at $z=0$, spanning an aperture $D$ with elements positioned at $x_n = (n - \frac{N+1}{2})d$, where \rev{$d = 0.49\lambda$} is the element spacing.
The $K$ users are distributed in the RNF region (Fresnel zone, $z < 2D^2/\lambda$) at coordinates $\mathbf{p}_k = (x_k, z_k)$.
\rev{Throughout the paper, $x$ denotes the \emph{continuous} transverse aperture coordinate with finite support $x\in[-D/2,\,D/2]$; continuous aperture envelopes are defined on this support and then sampled at the discrete element positions $x_n$. The analysis is two-dimensional and scalar in the $x$--$z$ plane: the transmitter is a single-axis ULA, all phase profiles (including the Airy profile below) are one-dimensional functions of $x$, and the extension to planar apertures and three-dimensional bending is left outside the scope of this paper.}

The complex baseband signal received by the $K$ users, $\mathbf{y} \in \mathbb{C}^{K \times 1}$, is given by:
\begin{equation}
    \mathbf{y} = \mathbf{H}_{\text{phys}} \mathbf{W}_{\text{RF}} \mathbf{W}_{\text{BB}} \mathbf{s} + \mathbf{n},
\end{equation}
where $\mathbf{H}_{\text{phys}} \in \mathbb{C}^{K \times N}$ is the physical channel matrix, $\mathbf{W}_{\text{RF}}$ and $\mathbf{W}_{\text{BB}}$ are the analog and digital precoders, respectively. $\mathbf{s} \in \mathbb{C}^{K \times 1}$ contains the normalized data symbols ($\mathbb{E}[\mathbf{s}\mathbf{s}^H] = \mathbf{I}_K$), and $\mathbf{n} \sim \mathcal{CN}(0, \sigma^2 \mathbf{I}_K)$ is the additive white Gaussian noise.

\subsection{Physical Channel Construction: Methodological Consistency}
\label{subsec:phys_channel}
Accurate modeling of RNF communications requires a rigorous wave-theoretic framework. In this work, we employ two complementary modeling approaches and explicitly establish their physical consistency.

\subsubsection{Baseline Green's Function Model (Blockage-Free)}
For LoS scenarios without obstacles, the channel response from the $n$-th antenna element to the $k$-th user is described by the free-space Green's function (cylindrical wave approximation):
\begin{equation}
    h_{k,n} = \frac{\lambda}{4\pi r_{k,n}} e^{-j k_0 r_{k,n}},
\end{equation}
where $k_0 = 2\pi/\lambda$ is the wavenumber and $r_{k,n} = \|\mathbf{p}_k - (x_n, 0)\|_2$. This captures the precise spherical wavefront curvature in the near field.

\subsubsection{Fresnel Diffraction Integration (With Blockage)}
\rev{To account for half-space blockage and the auto-bending propagation of Airy beams, we adopt a wave-optics approach based on the angular-spectrum method of Fourier optics \cite{GoodmanFourierOptics}. The propagation operator is defined explicitly, with its inverse transform shown, as $\mathcal{P}_{\Delta z}\{u\} = \mathcal{F}^{-1}\!\left\{ H(k_x,\Delta z)\, \mathcal{F}\{u\} \right\}$, where $\mathcal{F}$ is the transverse Fourier transform and $H(k_x,\Delta z)=\exp\!\big(-j\,\Delta z\sqrt{k_0^{2}-k_x^{2}}\big)$ is the exact-$k_z$ transfer function, consistent with the $e^{-jk_0 r}$ convention of (2). A \emph{knife edge} is defined as an opaque semi-infinite half-plane occupying $x \le \xobs$ in the transverse screen plane $z=\zobs$, with the complementary half-plane fully transmissive; the resulting edge diffraction is produced by the propagation operator itself. The field at the receiver depth $z_k$ is computed by propagating the aperture field to the screen, applying the mask, and propagating onward:}
\begin{equation}
    E(x, z_k) = \mathcal{P}_{\Delta z_2} \{ \mathcal{M}(x) \cdot \mathcal{P}_{\Delta z_1} \{ E_0(x) \} \},
\end{equation}
where $\mathcal{P}_{z}$ is the propagation operator defined above, and $\mathcal{M}(x)$ is the transmittance mask \rev{(unity on the open half-plane $x>\xobs$, zero on the obstacle)}. Here, ${\Delta z_1} = \zobs$ and ${\Delta z_2} = z_k - \zobs$ denote the propagation distances before and after the obstacle plane, respectively.

\rev{\textbf{Remark 1 (Model validity and numerical consistency):} The validity of this scalar beam-optics model is governed by the transverse spectral content of the field---equivalently, by the local phase-gradient (bending) angle---rather than by the propagation distance alone. The implemented operator uses the exact dispersion $k_z=\sqrt{k_0^2-k_x^2}$ and is therefore not restricted to the paraxial small-angle regime. For the representative scenes of Secs.~IV--V, the tangent-path angles span $4.6^{\circ}$--$11.2^{\circ}$, all sampling, window, and grid-convergence checks pass, and an independently implemented angular-spectrum reference reproduces the fields within $1.1\%$--$1.5\%$ and the resulting rates within $0.1\%$--$2.3\%$. The Green's-function model used for the blockage-free baseline (Sec.~III) and the angular-spectrum model used for the blockage studies (Secs.~IV--V) are numerically consistent where they overlap.}

\subsection{Hybrid Beamforming Strategy}
\label{subsec:beamforming}

\subsubsection{Analog Beamforming: Codebook and Optimization}
The analog precoder $\mathbf{W}_{\text{RF}} = [\mathbf{w}_1, \dots, \mathbf{w}_K] \in \mathbb{C}^{N \times K}$ \rev{connects each RF chain to one analog aperture beam, i.e., one column of $\mathbf{W}_{\text{RF}}$; no spherical-mode expansion is used anywhere in the model}. We focus on two strategies:

\begin{itemize}
    \item \textbf{Traditional Near-Field Focusing:} Matches the spherical phase of the user's location via phase conjugation:
    $w_{\text{trad}}^{(n)} = \frac{1}{\sqrt{N}} e^{j k_0 r_{k,n}}$.
    
    \item \textbf{Proposed Airy Beamforming:} We impart a cubic phase modulation derived from the paraxial finite-energy Airy beam formulation \cite{ref10}:
    \begin{equation}
        \phi_{\text{Airy}}(x_n) = \frac{k_0}{2F}x_n^2 - k_0 \sin(\theta) x_n + \frac{2\pi}{3\lambda} B \left(\frac{x_n}{F}\right)^3.
        \label{eq:airy_phase}
    \end{equation}
    \rev{Here, $\theta$ is the \emph{launch angle}: the linear phase tilt sets the direction in which the beam leaves the aperture. $F$ is a \emph{virtual focal length}: it scales the quadratic term and sets the longitudinal convergence scale of the envelope; the beam does not actually focus at $z=F$, hence ``virtual.'' $B$ is a dimensionless \emph{signed curvature coefficient}: it scales the cubic term, sets the transverse acceleration of the intensity ridge, and thereby controls the clearance with which the main lobe passes the obstacle edge. No single parameter produces the blockage gain by itself (Sec.~V).
    \textbf{Optimization Insight:} Unlike static parameter settings, the tuple $(B, F, \theta)$ provides degrees of freedom to manipulate the beam trajectory. In multi-user scenarios (Sec.~V), these parameters are tuned to improve the whitened separability of the effective channel and the service delivered to the shadowed user, while the final inter-user cancellation is performed by the digital stage. As the per-user field evidence in Sec.~V shows, the optimized beam can \emph{raise} raw cross-user leakage while still improving the end-to-end channel: the analog stage conditions the channel rather than steering nulls.}
\end{itemize}

\subsubsection{Digital Precoding (Zero-Forcing)}
Defining the effective channel as $\mathbf{H}_{\text{eff}} = \mathbf{H}_{\text{phys}} \mathbf{W}_{\text{RF}}$, we employ Zero-Forcing (ZF) to decouple user streams:
\begin{equation}
    \mathbf{W}_{\text{BB}} = \alpha \mathbf{H}_{\text{eff}}^H (\mathbf{H}_{\text{eff}}\mathbf{H}_{\text{eff}}^H)^{-1},
    \label{eq:ZF_precoding}
\end{equation}
where $\alpha$ is the power normalization factor such that $\|\mathbf{W}_{\text{RF}}\mathbf{W}_{\text{BB}}\|_F^2 = P_{\text{tx}}$.
\rev{Since $\|\mathbf{W}_{\text{RF}}\mathbf{W}_{\text{BB}}\|_F^2=\operatorname{tr}\!\big(\mathbf{W}_{\text{BB}}^H\mathbf{G}\,\mathbf{W}_{\text{BB}}\big)$ with the RF Gram matrix $\mathbf{G}=\mathbf{W}_{\text{RF}}^H\mathbf{W}_{\text{RF}}$, this constraint charges the design for the true radiated power even when the analog beams are strongly non-orthogonal (their measured correlation reaches $0.86$ in the mixed scene of Sec.~V). Normalizing only $\mathbf{W}_{\text{BB}}$ would implicitly treat the analog beams as orthogonal and undercharge the radiated power; every result in this paper therefore enforces the total-power constraint on the full product $\mathbf{W}_{\text{RF}}\mathbf{W}_{\text{BB}}$.}

\subsection{Performance Metrics and Analysis}
To evaluate the system robustness under blockage, we utilize the following metrics.

\subsubsection{Whitened Channel Conditioning}
\rev{The condition number of the effective channel, $\kappa(\mathbf{H}_{\text{eff}}) = \sigma_{\max} / \sigma_{\min}$, quantifies the orthogonality of the user subspaces; a collapse of $\sigma_{\min}$ under blockage is what defeats ZF. However, raw $\kappa(\mathbf{H}_{\text{eff}})$ is not invariant across schemes whose analog beams are non-orthogonal: it mixes the conditioning of the propagation channel with that of the beam basis, and a scheme could ``improve'' it merely by choosing more mutually orthogonal beams. Under the total-power constraint above, the substitution $\mathbf{W}_{\text{BB}}=\mathbf{G}^{-1/2}\bar{\mathbf{W}}$ shows that the system is exactly equivalent to one with the \emph{whitened} channel $\bar{\mathbf{H}}=\mathbf{H}_{\text{eff}}\,\mathbf{G}^{-1/2}$ under a plain Frobenius power budget. All cross-scheme separability statements in this paper therefore use $\bar{\mathbf{H}}$, and because no single scalar determines ZF noise enhancement, we always report three quantities jointly: the condition number $\kappa(\bar{\mathbf{H}})$, the weak-mode strength $\sigma_{\min}(\bar{\mathbf{H}})$, and the ZF power cost $\sum_i \sigma_i^{-2}(\bar{\mathbf{H}})$.}

\subsubsection{Signal-to-Interference-plus-Noise Ratio (SINR)}
With the ZF precoding design described in~(\ref{eq:ZF_precoding}), the effective channel is forced to be diagonal: $\mathbf{H}_{\text{eff}}\mathbf{W}_{\text{BB}} \approx \alpha \mathbf{I}_K$. Consequently, the received SINR for the $k$-th user is given by:
\begin{equation}
    \text{SINR}_k = \frac{|\alpha|^2}{\sigma^2}.
    \label{eq:sinr_equal}
\end{equation}

\textbf{Remark 2 (Equal SINR Implications):} 

The SINR expression in~(\ref{eq:sinr_equal}) corresponds to the \emph{post-precoding effective SINR} under ZF baseband precoding. Specifically, with the ZF design in~(\ref{eq:ZF_precoding}), the equivalent channel satisfies
\begin{equation}
    \mathbf{H}_{\mathrm{eff}}\mathbf{W}_{\mathrm{BB}} = \alpha\,\mathbf{I}_K,
    \label{eq:sinr_normalized}
\end{equation}
which implies that all users experience the same effective gain $|\alpha|^2$ and hence identical SINR values. This does \emph{not} imply that the underlying physical channels of shadowed and line-of-sight (LoS) users are identical. Instead, ZF enforces stream decoupling and equalization, while the channel disparity is absorbed into the global scaling factor $\alpha$. When $\mathbf{H}_{\mathrm{eff}}$ becomes ill-conditioned (e.g., under severe blockage), $\alpha$ sharply decreases due to noise amplification, leading to a simultaneous SINR degradation for all users.

We adopt this equalized ZF formulation to explicitly expose the effect of singular-value collapse and rank recovery in near-field multi-user channels. Alternative designs such as weighted ZF or explicit power allocation can yield unequal per-user SINR, but are outside the scope of this work.
\rev{In addition to the sum rate $R_{\text{sum}}=\sum_k \log_2(1+\mathrm{SINR}_k)$, we report the \emph{minimum user rate} wherever multiple users are served, so that a large bright-user rate cannot conceal a shadow-user outage.}

\section{Baseline RNF Multiuser System with Conventional Focusing}
\label{sec:baseline}

This section establishes the performance baseline for a radiative near-field (RNF) multiuser (MU) system utilizing conventional focused beams in a blockage-free environment. By characterizing the system's intrinsic spatial resolution using the rigorous Green's function model (defined in Section~\ref{subsec:phys_channel}), we provide a theoretical upper bound. This benchmark is crucial for quantifying the severity of the blockage impairments discussed later in Section~IV.

\subsection{Simulation Geometry and Parameters}

\begin{figure*}[!t]
    \centering
    \includegraphics[width=\textwidth]{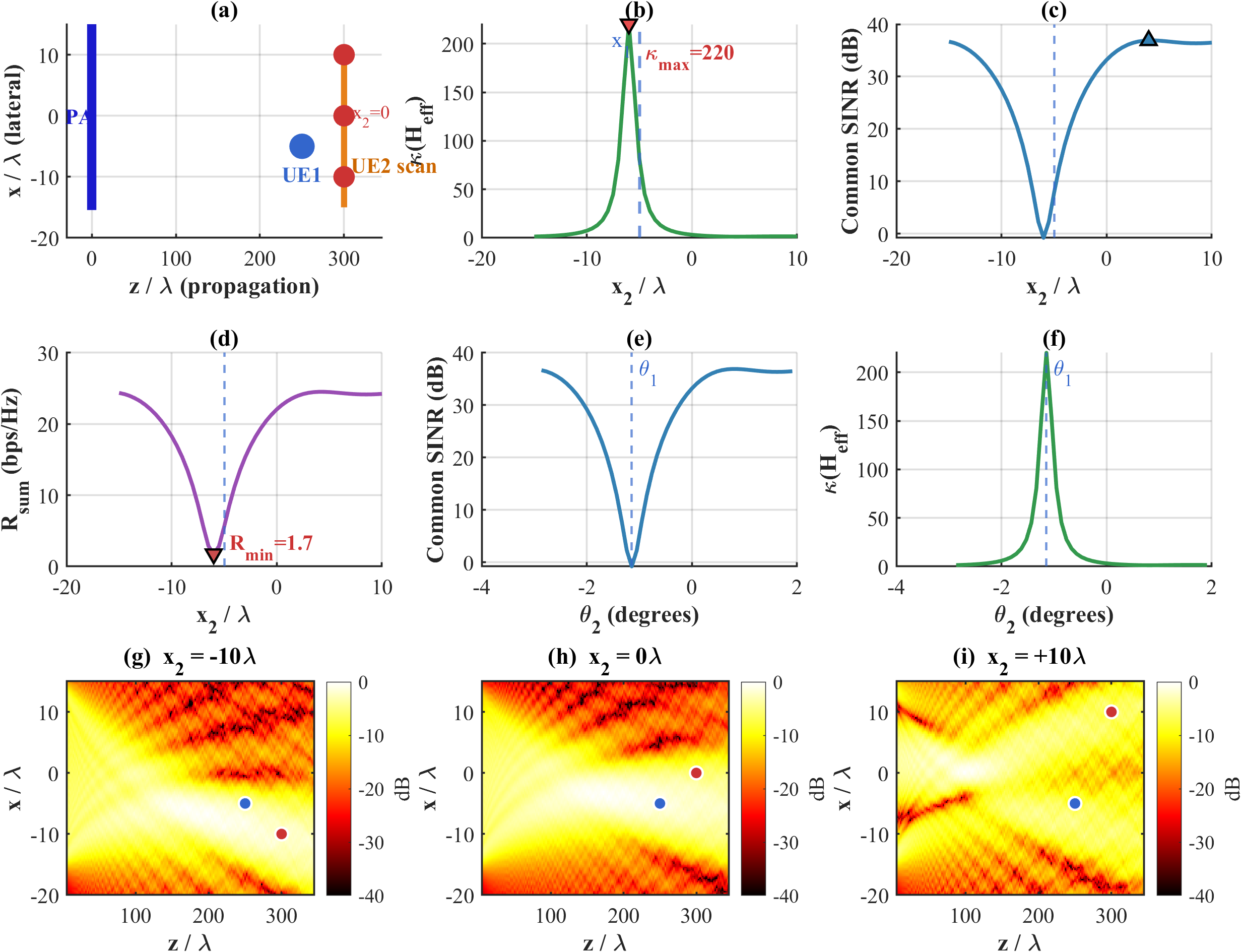}
    \caption{Baseline performance of the RNF-MU system with conventional focusing in free space.
    (a) System Geometry\rev{ (blockage-free; the obstacle plane of Sec.~IV plays no role here and is no longer drawn)}: UE-1 fixed at $(-5\lambda, 250\lambda)$ and UE-2 scanning at $z=300\lambda$.
    (b) Condition Number $\kappa(\mathbf{H}_{\text{eff}})$: A sharp singularity (\rev{$\kappa_{\max} \approx 220$}) occurs at $x_2 = -6\lambda$, corresponding to angular alignment with UE-1.
    (c) Common SINR: Equalized SINR profile showing degradation at the angular collision.
    (d) Sum-Rate: Minimum throughput observed at the singularity point.
    (e)--(f) Angular Analysis: SINR and $\kappa$ plotted against UE-2's angle $\theta_2$, confirming the overlap occurs at $\theta_2 \approx \theta_1 \approx -1.15^\circ$.
    (g)--(i) Field Intensity Distributions: Normalized field power $|E(x,z)|^2$ for UE-2 located at $x_2 = -10\lambda$, $0\lambda$, and $+10\lambda$, respectively, visualizing the distinct focal spots.}
    \label{fig:baseline}
\end{figure*}

We adopt the 2D simulation geometry illustrated in Fig.~\ref{fig:baseline}(a). The BS employs a large-scale ULA with $N=64$ elements and half-wavelength spacing ($d = 0.49\lambda$), operating at $f_c = 28$ GHz. This configuration yields a Fraunhofer distance of $R_{\text{FF}} \approx 1966\lambda$, placing the users deeply within the radiative near-field.

To ensure continuity with the blockage scenarios in subsequent sections, the user deployment is configured as follows:
\begin{itemize}
    \item \textbf{UE-1 (Target User):} Fixed at coordinates $\mathbf{p}_1 = (-5\lambda, 250\lambda)$.
    \textit{Remark:} While this location is Line-of-Sight (LoS) in this baseline section, it is strategically chosen to fall within the ``geometric shadow'' region ($x < 0$) once the half-space obstacle is introduced in Section~IV.
    \item \textbf{UE-2 (Interfering User):} Located at a different depth $z_2 = 300\lambda$. This user scans laterally along the $x$-axis from $x_2 = -15\lambda$ (deep shadow side) to $+10\lambda$ (bright side), crossing the angular line-of-sight of UE-1.
\end{itemize}

The system supports $N_s=2$ spatial streams with a normalized transmit power $P_{\text{tx}}=1$ \rev{and, for this free-space baseline study, a normalized noise floor of $N_0=10^{-3}$, which provides a reference SNR of approximately $20$~dB. The quantitative blockage campaigns of Secs.~IV-D and V declare a separately calibrated contract ($N_0=5.826\times 10^{-4}$, fixed once by an exact $20$-dB free-space matched-beam reference); each figure states the convention under which it was produced.}

\subsection{Conventional Beamforming and Evaluation}

\subsubsection{Green's Function Channel Construction}
In this blockage-free baseline, the physical channel vector $\mathbf{h}_k \in \mathbb{C}^{N \times 1}$ is generated directly using the free-space Green's function model detailed in Section~\ref{subsec:phys_channel}. This ensures that the baseline evaluation captures the exact spherical wavefront curvature and path loss inherent to the RNF region, serving as a rigorous ``ground truth'' for comparison with the diffraction-based models in subsequent sections.

\subsubsection{Hybrid Precoding Strategy}
The analog precoder $\mathbf{W}_{\text{RF}}$ employs Traditional Near-Field Focusing weights (phase conjugation) perfectly matched to each user's location.

For the digital baseband precoder $\mathbf{W}_{\text{BB}}$, we employ Regularized Zero-Forcing (RZF) to mitigate multi-user interference while maintaining numerical stability near channel singularities. The unnormalized precoding matrix is given by:
\begin{equation}
    \tilde{\mathbf{W}}_{\text{BB}} = \mathbf{H}_{\text{eff}}^H \left( \mathbf{H}_{\text{eff}} \mathbf{H}_{\text{eff}}^H + \epsilon \mathbf{I}_K \right)^{-1},
\end{equation}
where $\epsilon \approx 10^{-10}$ is a regularization factor. The final precoder is power-normalized as $\mathbf{W}_{\text{BB}} = \alpha \tilde{\mathbf{W}}_{\text{BB}}$ to satisfy $\|\mathbf{W}_{\text{RF}}\mathbf{W}_{\text{BB}}\|_F^2 = P_{\text{tx}}$.

\subsection{Performance Analysis: Resolution and Conditionality}

Fig.~\ref{fig:baseline} presents the comprehensive performance analysis. 

\subsubsection{Condition Number and Angular Singularity}
Fig.~\ref{fig:baseline}(b) plots the condition number $\kappa(\mathbf{H}_{\text{eff}})$ against the lateral position of UE-2. \rev{(In this free-space baseline both analog beams are focused beams of the same family, so raw $\kappa(\mathbf{H}_{\text{eff}})$ serves as a within-scheme singularity indicator; every cross-scheme comparison from Sec.~IV onward uses the whitened metrics of Sec.~II-D.)} A critical observation is that the peak condition number (\rev{$\kappa \approx 220$}) occurs at $x_2 = -6\lambda$, not at $x_2 = -5\lambda$.
This shift is physically significant:
\begin{itemize}
    \item UE-1 is at $(-5\lambda, 250\lambda)$, corresponding to an angle $\theta_1 \approx \text{atan}(-5/250) \approx -1.15^\circ$.
    \item The singularity occurs when UE-2 (at $z_2=300\lambda$) aligns with this angle: $x_2 = 300\lambda \cdot \tan(\theta_1) = -6\lambda$.
\end{itemize}
Thus the collision occurs when UE-2 shares the same ray angle with UE-1, not when $x_2$ equals $x_1$. This indicates that for the given aperture ($D \approx 31\lambda$) and depth separation ($50\lambda$), \textit{traditional focusing relies primarily on angular separation} to maintain orthogonality. The longitudinal resolution is insufficient to decorrelate users located on the same ray, leading to a ``co-linear singularity.''

\subsubsection{Spectral Efficiency}
Reflecting the conditioning analysis, the Sum-Rate (Fig.~\ref{fig:baseline}(d)) and Common SINR (Fig.~\ref{fig:baseline}(c)) exhibit a deep notch centered at the angular collision point ($x_2 = -6\lambda$). Outside this narrow angular window (approx. $\pm 2^\circ$, as seen in Fig.~\ref{fig:baseline}(e)), the SINR recovers rapidly to its maximum plateau. This confirms that RNF MU-MIMO with traditional beams is highly effective but remains vulnerable to ``flashlight effect'' interference when users share the same Line-of-Sight angle.

\emph{Summary:} The baseline analysis demonstrates that while traditional beams offer high gain, they suffer from rank collapse under angular alignment, even with RZF precoding. In the next section, we will see how a half-space obstacle exacerbates this by creating a ``permanent'' singularity zone (shadow) where traditional beams fail completely, motivating the need for Airy beamforming.

\section{Half-Space Blockage: Double-Shadow Users and Airy-Geo Analog Beamforming}
\label{sec:blockage}

Having established the baseline performance in free space in Section~III, we now investigate the critical scenario of half-space blockage. This section reveals the vulnerability of traditional near-field focusing when Line-of-Sight (LoS) paths are obstructed and demonstrates how the proposed Airy beamforming strategy---leveraging its auto-bending property---restores connectivity and recovers the channel rank in the geometric shadow region.

\subsection{Half-Space Blockage Scenario and Singular Value Collapse}

We consider a scenario where a semi-infinite obstacle (e.g., a wall or pillar) creates a half-space blockage. As illustrated in Fig.~\ref{fig:blockage_results}(a), a knife-edge obstacle is located at depth $\zobs = 150\lambda$, blocking the entire region $x \le \xobs = 0$.

To rigorously evaluate the system's resilience, we focus on a challenging ``Double Shadow'' configuration where both users are located deep within the geometric shadow zone. To ensure a fair comparison with the baseline in Section~III:
\begin{itemize}
    \item \textbf{UE-1 (Target):} Remains fixed at $(x_1, z_1) = (-5\lambda, 250\lambda)$, consistent with the baseline geometry (except now obstructed).
    \item \textbf{UE-2 (Scanning):} Is located at depth $z_2 = 300\lambda$ but is now scanned laterally from $x_2 = -15\lambda$ (deep shadow) to $x_2 = -1\lambda$ (near the diffraction edge). This range is specifically chosen to investigate performance deep inside the shadow where traditional links fail.
\end{itemize}

In this setup, the direct LoS paths between the array and the users are geometrically severed. For traditional near-field focusing, which relies on linear propagation, energy transfer is restricted to weak diffraction components radiating from the obstacle's edge. This leads to a catastrophic \textbf{singular value collapse} of the effective multi-user channel $\Heff$. The channel matrix becomes ill-conditioned ($\kappa(\Heff) \to \infty$), and the ZF precoder fails to invert the channel without severely amplifying noise, resulting in a communication outage.

\subsection{Proposed Airy-Geo Analog Beamforming Strategy}

To overcome the shadow, we propose an \textbf{``Airy-Geo''} analog beamforming strategy. Unlike traditional beams that focus energy on a straight line, Airy beams follow a curved parabolic trajectory, enabling them to ``ride'' the diffraction edge and accelerate into the shadow zone.

\subsubsection{Fixed Geometry-Based Codebook Design}
\begin{figure*}[!t]
    \centering
    \includegraphics[width=\textwidth]{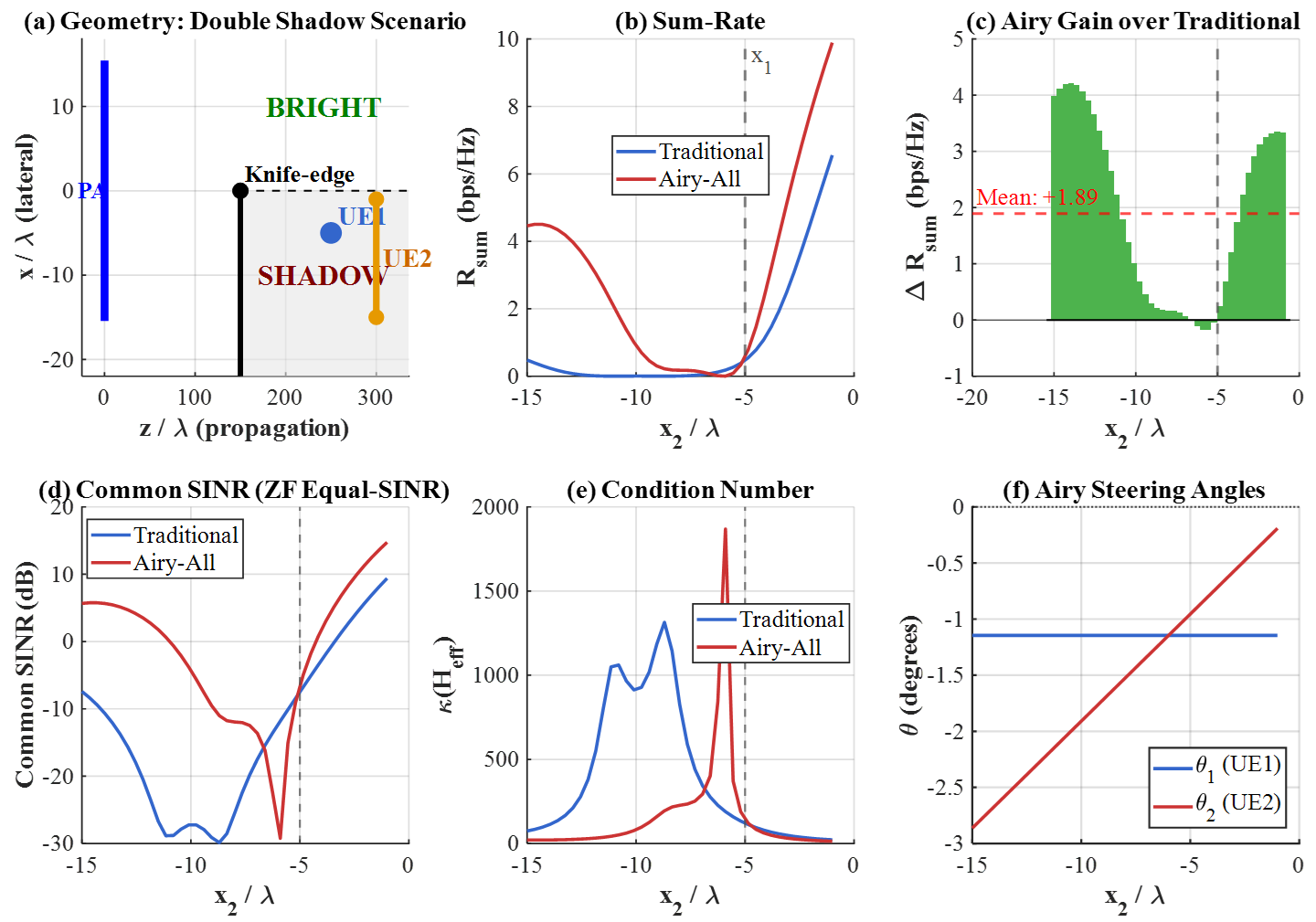}
    

    \caption{Performance comparison in the double-shadow scenario ($x_1=-5\lambda$, fixed; $x_2$ scanning inside shadow). \rev{``Airy-All'' denotes the scheme in which every analog beam is drawn from the Airy codebook, and ``Trad-All'' the all-focusing counterpart.} (a) System geometry showing the knife-edge blockage at $z=150\lambda$. (b) Sum-rate performance: Airy beams (red) maintain a robust rate floor in the deep shadow where traditional beams (blue) suffer outage. (c) Sum-rate gain of Airy over traditional focusing, showing gains up to 4 bps/Hz. (d) Common SINR: Airy beams provide a usable SINR floor despite interference, whereas traditional SINR collapses. (e) Condition number: Note that while Airy has a higher $\kappa$ near collision, it provides non-zero singular values, enabling communication. (f) Steering angles used for the Airy beams, following simple geometric logic.}
    \label{fig:blockage_results}
\end{figure*}

A key feature of our proposal is the use of a simple, geometry-based analog codebook. We deliberately avoid complex, per-user optimization of the beam parameters to demonstrate the intrinsic robustness of the Airy wavepacket. The design follows two principles:

\begin{enumerate}
    \item \textbf{Physics-Informed Parameter Selection:} We fix the curvature parameters $(B, F)$ for all users to maximize the ``edge-riding'' effect.
    \begin{itemize}
        \item The focal length is set to $F \approx 163\lambda$ ($\approx$ 1.75 m at 28 GHz, $\lambda \approx$ 10.7 mm), slightly beyond the obstacle depth ($\zobs = 150\lambda$). This ensures that the beam's main lobe is fully formed and begins its transverse acceleration exactly as it passes the diffraction edge.
        \item The bending coefficient is set to $B = -25$. This relatively strong negative value imparts a significant transverse acceleration to the main lobe, bending it towards the \rev{negative} $x$-direction (into the shadow) as it propagates. \rev{(This is a fixed demonstration value for the codebook of this section; the jointly optimized parameters of Sec.~V are different and should not be conflated with it.)}
    \end{itemize}
    
    \item \textbf{Geometric Steering:} For each user $k$, the initial launch angle $\theta_k$ is set simply to the geometric angle $\theta_k = \operatorname{atan2}(x_k, z_k)$, as plotted in Fig.~\ref{fig:blockage_results}(f). This confirms that our strategy relies purely on geometry without requiring complex channel state information for analog beam selection.
\end{enumerate}

\textbf{Remark \rev{3} (Sub-optimality of Geometric Steering):} It is important to acknowledge that in a blockage scenario, pointing the beam directly at the user ($\theta_k$) is not strictly optimal. The optimal strategy would likely involve aiming at a virtual source near the diffraction edge to maximize diffracted energy flux. However, we adopt this sub-optimal, geometry-based steering to highlight that even a constrained Airy design significantly outperforms traditional focusing due to its inherent self-healing physics.

\begin{figure*}[!t]
    \centering
    \includegraphics[width=0.95\linewidth]{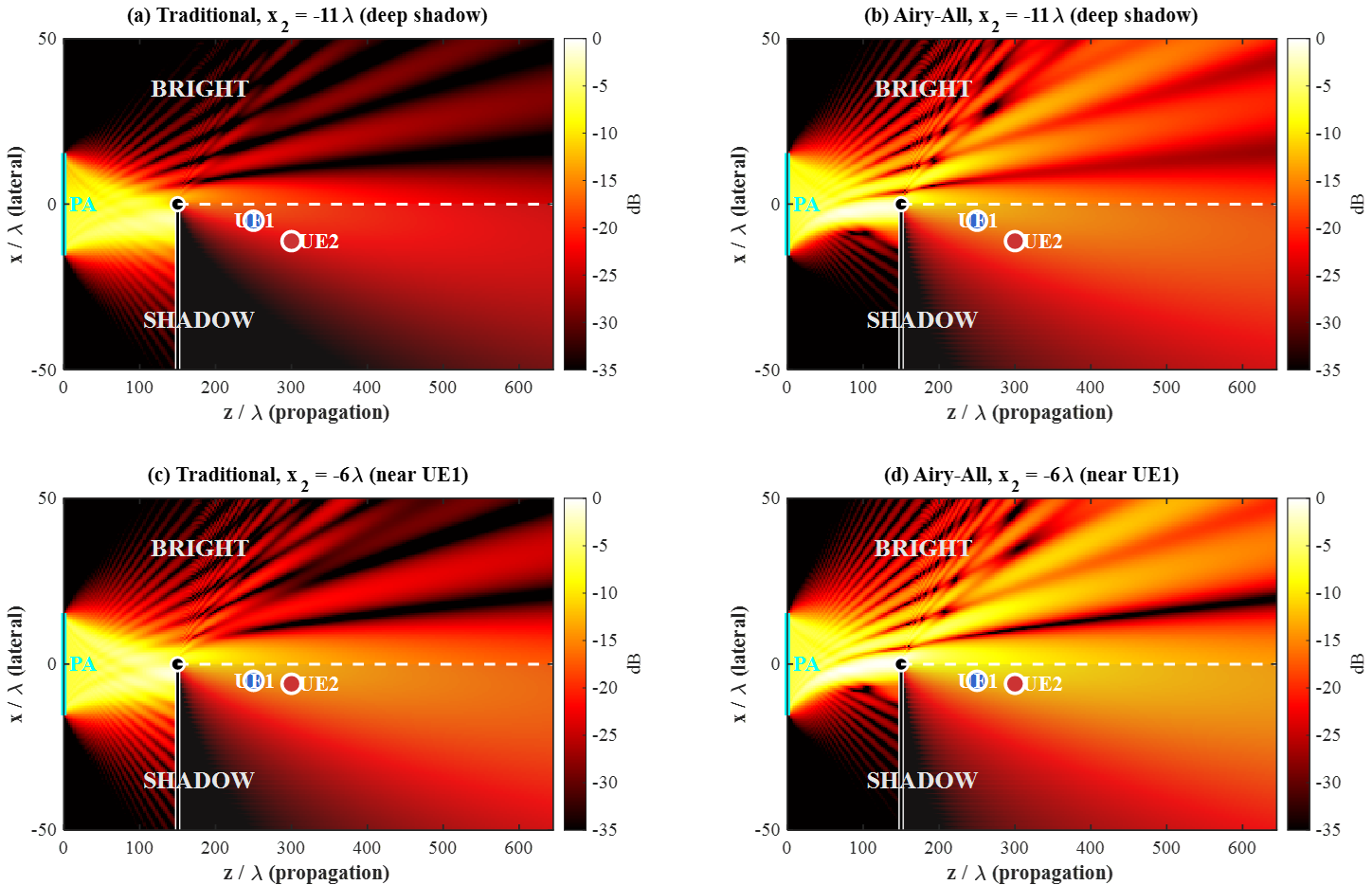}

    \caption{Normalized field intensity distributions $|E(x,z)|^2$ (dB). Top Row (a, b): Deep shadow scenario ($x_2 = -11\lambda$). Traditional beams (a) are severed, while Airy beams (b) bend significantly to serve the distant user. Bottom Row (c, d): Angular collision scenario ($x_2 \approx x_1 = -5\lambda$). Traditional beams (c) fail completely, while Airy beams (d) successfully deliver power to both co-located users despite the spatial overlap. Parameters: $B=-25, F \approx 163\lambda$.}
    \label{fig:field_distribution}
\end{figure*}

\subsection{Performance Evaluation and Analysis}

We evaluate the performance using the rigorous Fresnel diffraction channel model described in Section~II. To reflect a realistic low-SNR regime typical of non-LoS scenarios, the noise power is set to $N_0 = 10^{-3}$ (normalized). The results are summarized in Fig.~\ref{fig:blockage_results} and visualized in Fig.~\ref{fig:field_distribution}.

\subsubsection{SINR Resilience in Deep Shadow}
Fig.~\ref{fig:blockage_results}(d) compares the Common SINR of the Traditional and Airy-Geo schemes.
\begin{itemize}
    \item \emph{Traditional Failure:} As expected, the traditional focusing scheme (blue curve) experiences a severe outage in the deep shadow region ($x_2 < -6\lambda$). The SINR drops significantly below 0 dB, indicating that the diffracted signal power is buried in the noise. This confirms that linear beams cannot effectively deliver energy behind the half-space blockage.
    \item \emph{Airy Resilience:} In contrast, the Airy scheme (red curve) maintains a usable SINR level \rev{over most of the shadow region, although the deepest portion of the scan can still fall below $0$~dB; per-user service is therefore reported under a declared service floor in Secs.~IV-D and V.} This SINR resilience is critical: it implies that the Airy beam successfully establishes a viable communication link where traditional methods fail completely. While the absolute sum-rate (Fig.~\ref{fig:blockage_results}(b)) is naturally constrained by the high noise floor, the relative advantage of maintaining connectivity is significant.
\end{itemize}

\subsubsection{Mechanism of Recovery: Active Bending}
The physical mechanism behind this gain is visualized in Fig.~\ref{fig:field_distribution}.
\begin{itemize}
    \item \emph{Deep Shadow Recovery (Top Row):} Fig.~\ref{fig:field_distribution}(a) and (b) correspond to a large separation ($x_2 = -11\lambda$). Traditional beams (a) are sharply truncated, creating a dark triangular shadow zone. In contrast, Airy beams (b) exhibit a clear curved trajectory. The main lobe accelerates transversely, effectively ``turning the corner'' of the obstacle to illuminate the deep shadow user UE-2.
    \item \emph{Recovery under Angular Collision (Bottom Row):} Fig.~\ref{fig:field_distribution}(c) and (d) depict the scenario where UE-2 is close to UE-1 ($x_2 \approx -5\lambda$). Here, both users are shadowed. While traditional beams (c) result in a complete double outage, the Airy beams (d) manage to illuminate both locations. Although the beams spatially overlap near the obstacle edge, they retain sufficient distinctness in the depth domain to allow the ZF precoder to separate the streams.
\end{itemize}

\subsubsection{The Trade-off: Power vs. Orthogonality}
In Fig.~\ref{fig:blockage_results}(e), the condition number $\kappa(\Heff)$ of the traditional near-field focusing scheme appears numerically smaller than that of the Airy-based scheme in the deep-shadow regime. This observation should be interpreted with care. For the traditional design, severe blockage drives \emph{both} the largest and smallest singular values of $\Heff$ close to zero, i.e., $\sigma_{\max}\approx 0$ and $\sigma_{\min}\approx 0$. As a result, the numerical value of $\kappa=\sigma_{\max}/\sigma_{\min}$ becomes highly sensitive to numerical truncation and noise-floor effects, and does not reflect a usable channel. Indeed, the corresponding absolute channel gain is extremely small, leading to SINR outage despite a seemingly moderate condition number.

In contrast, the Airy-based transmission significantly elevates $\sigma_{\min}$ above the noise floor by delivering non-negligible diffracted energy into the shadow region. While diffraction-induced beam spreading may increase channel correlation and hence yield a larger $\kappa$, the resulting effective channel is \emph{operational}, supporting reliable communication with a non-zero SINR. Therefore, in deep-shadow scenarios, absolute channel gain and the recovery of the smallest singular value are more relevant performance indicators than the condition number alone. This distinction explains why the Airy-based scheme achieves superior sum-rate performance even when its condition number is numerically larger.

However, the SINR results in Fig.~\ref{fig:blockage_results}(d) conclusively demonstrate that \emph{recovering signal power is paramount} in this noise-limited regime; the system achieves a viable link despite the increased interference, which is subsequently managed by the ZF precoder.

\subsection{Blockage-Geometry Capability, Whitened Separability, and Applicability Boundary}
\label{subsec:breadth}
\begin{figure*}[!t]
    \centering
    \begin{tikzpicture}[x=1cm,y=1cm,
        ap/.style={line width=2.6pt},
        obst/.style={fill=black!75,draw=none},
        ray/.style={densely dashed,black!50,line width=0.5pt},
        swp/.style={{Stealth[length=1.5mm]}-{Stealth[length=1.5mm]},black!60,line width=0.6pt},
        ue1/.style={circle,draw=black,fill=cyan!60,inner sep=1.6pt},
        ue2/.style={rectangle,draw=black,fill=yellow!60,inner sep=1.9pt},
        lab/.style={font=\scriptsize,align=center}]
      \node[font=\bfseries] at (-0.5,2.75) {(a)};
      \begin{scope}[shift={(0,0)}]
        \draw[ap] (0,0.35) -- (0,2.45); \node[lab] at (0,0.08) {BS};
        \fill[obst] (1.75,0.30) rectangle (1.95,1.45);
        \draw[ray] (0,1.5) -- (1.85,1.45) -- (3.2,0.78);
        \draw[ray] (0,1.7) -- (3.2,2.12);
        \node[ue1] at (3.3,0.75) {}; \node[ue2] at (3.3,2.15) {};
        \draw[swp] (2.3,0.5) -- (2.3,2.2);
        \node[lab] at (1.85,-0.42) {half-space knife edge\\ fraction / transmittance /\\ depth swept};
      \end{scope}
      \begin{scope}[shift={(4.55,0)}]
        \draw[ap] (0,0.35) -- (0,2.45); \node[lab] at (0,0.08) {BS};
        \fill[obst] (1.75,0.95) rectangle (1.95,1.85);
        \draw[ray] (0,1.15) -- (1.85,0.95) -- (3.2,0.78);
        \draw[ray] (0,1.75) -- (1.85,1.85) -- (3.2,2.12);
        \node[ue1] at (3.3,0.75) {}; \node[ue2] at (3.3,2.15) {};
        \draw[swp] (2.3,0.95) -- (2.3,1.85);
        \node[lab] at (1.85,-0.42) {finite strip (pillar)\\ width / center /\\ transmittance swept};
      \end{scope}
      \begin{scope}[shift={(9.1,0)}]
        \draw[ap] (0,0.35) -- (0,2.45); \node[lab] at (0,0.08) {BS};
        \fill[obst] (1.30,0.30) rectangle (1.50,1.30);
        \fill[obst] (2.10,0.30) rectangle (2.30,1.55);
        \draw[black!60,line width=0.6pt,-{Stealth[length=1.5mm]}] (1.40,1.62) -- ++(15:0.45);
        \draw[black!60,line width=0.6pt,-{Stealth[length=1.5mm]}] (2.20,1.87) -- ++(-15:0.45);
        \draw[ray] (0,1.45) -- (1.4,1.3) -- (2.2,1.55) -- (3.15,0.85);
        \draw[ray] (0,1.8) -- (3.15,2.3);
        \node[ue1] at (3.25,0.62) {}; \node[ue1] at (3.4,1.02) {};
        \node[ue2] at (3.25,1.92) {}; \node[ue2] at (3.4,2.32) {};
        \node[lab] at (1.85,-0.42) {two blockers,\\ edge tilt $\pm 15^{\circ}$,\\ $K=2$--$4$ users};
      \end{scope}
      \begin{scope}[shift={(13.65,0)}]
        \draw[ap] (0,0.35) -- (0,2.45); \node[lab] at (0,0.08) {BS};
        \fill[obst] (1.75,0.30) rectangle (1.95,2.45);
        \draw[ray] (0,1.2) -- (1.75,1.05);
        \draw[ray] (0,1.6) -- (1.75,1.75);
        \node[red,font=\normalsize\bfseries] at (2.35,1.4) {$\times$};
        \node[ue1] at (3.3,0.75) {}; \node[ue2] at (3.3,2.15) {};
        \node[lab] at (1.85,-0.42) {fully opaque plane\\ no edge $\Rightarrow$ outage\\ (invariant, excluded)};
      \end{scope}
    \end{tikzpicture}\\[4pt]
    \includegraphics[width=.92\textwidth]{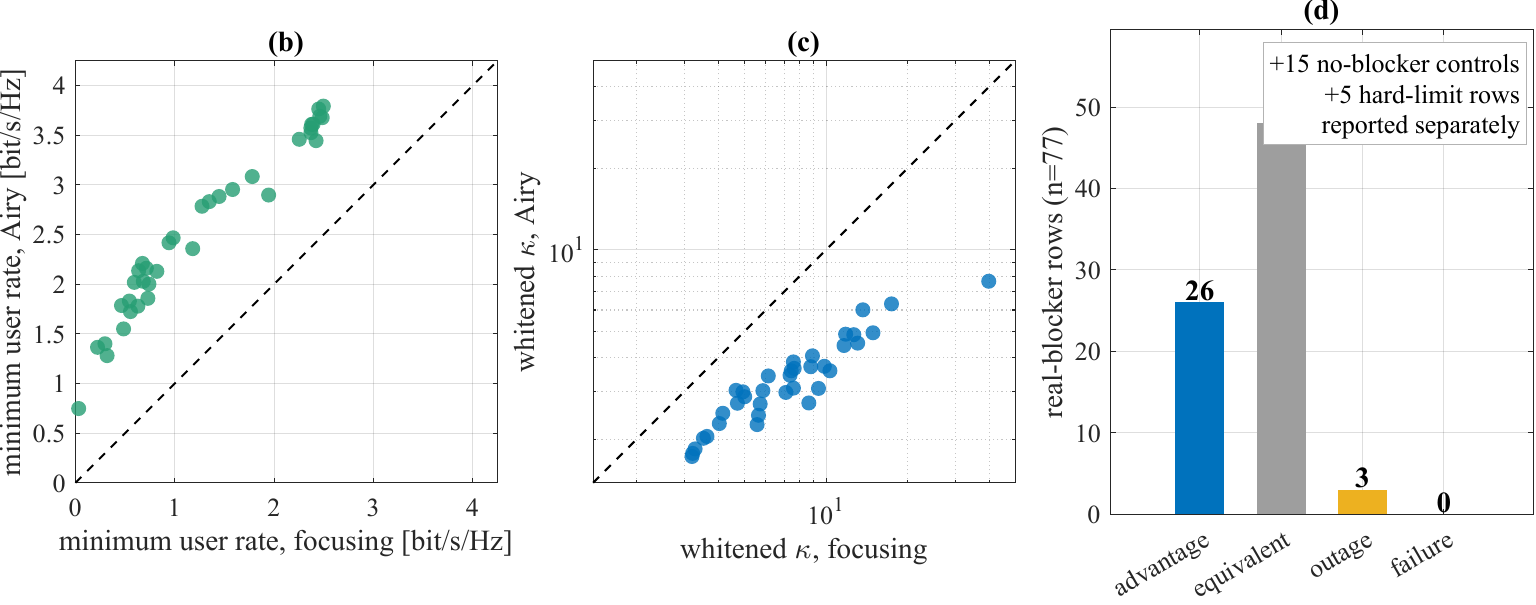}
    \caption{\rev{Blockage-geometry capability under the declared contract. (a)~Obstacle families of the pre-specified campaign (circles: shadowed users; squares: bright users): a semi-infinite knife edge with swept blocked fraction ($0$--$1$), transmittance ($0$--$1$), and depth ($0.4$--$0.8\,z_1$); a finite opaque strip with swept width ($0.10$--$0.75\,D$), center ($\pm 0.25\,D$), and transmittance ($0$--$0.6$); two-blocker scenes with effective edge tilts of $\pm 15^{\circ}$ serving $K=2$--$4$ users; and the fully opaque full plane, which defines the no-edge outage invariant and is excluded from the statistics. (b)~Minimum user rate across the $36$ blocked-user maps: improved in $36$ of $36$ (median $0.96\to 2.39$~bit/s/Hz). (c)~Whitened conditioning: improved in $36$ of $36$ (median $7.40\to 3.09$). (d)~Stratified outcome of the $77$ real-blocker rows: $26$ advantage / $48$ equivalent / $3$ outage / $0$ failure (equivalence band and service floor $0.05$~bit/s/Hz), with $15$ no-blocker controls and $5$ hard-limit rows reported separately.}}
    \label{fig:breadth}
\end{figure*}

\rev{The preceding subsections demonstrate the mechanism on fixed geometries; this subsection tests breadth with a pre-specified campaign over obstacle configurations, all evaluated under one executable contract: the exact-$k_z$ angular-spectrum propagator of Sec.~II, the sampled-ULA Gram matrix $\mathbf{G}=\mathbf{V}^H\mathbf{V}$ built from the discretized analog beams, the total-power constraint of Sec.~II-C, and a noise floor $N_0=5.826\times 10^{-4}$ calibrated once to an exact $20$-dB free-space matched-beam reference. Fig.~\ref{fig:breadth}(a) shows the obstacle families swept: a semi-infinite knife edge (blocked fraction $0$--$1$, transmittance $0$--$1$, depth $0.4$--$0.8\,z_1$), a finite opaque strip (width $0.10$--$0.75\,D$, center $\pm 0.25\,D$, transmittance $0$--$0.6$), two-blocker scenes with effective edge tilts of $\pm 15^{\circ}$ and $K=2/3/4$ shadow/bright user mixes, and a fully opaque full plane. Edge tilt is an effective proxy within the two-dimensional $x$--$z$ model; arbitrary three-dimensional wall orientation is outside the scope.}

\rev{The core multiuser result is Fig.~\ref{fig:breadth}(b)--(c): on the $36$ maps in which a user is actually blocked, the optimized Airy family improves every separability measure of Sec.~II-D in every scene. The whitened condition number improves in $36$ of $36$ (median $7.40\to 3.09$), the whitened $\sigma_{\min}$ rises in $36$ of $36$ (median $0.0237\to 0.0519$), and, most directly, the minimum user rate rises in $36$ of $36$, with the median worst-user rate more than doubling ($0.96\to 2.39$~bit/s/Hz). All users clear $0$-dB post-ZF SINR in $35$ of $36$ blocked-user scenes, against $17$ of $36$ under focusing.}

\rev{Fig.~\ref{fig:breadth}(d) places this across the full $77$-row real-blocker sweep: $26$ rows favor the optimized Airy family beyond the $0.05$-bit/s/Hz equivalence band, $48$ are equivalent, $3$ are outages in which no scheme reaches the service floor, and none favors focusing; the $15$ no-blocker controls and $5$ physical/search hard-limit rows are reported separately. The zero count on the focusing side is structural rather than remarkable: pure focusing is the $B=0$ member of the searched family, so the family cannot lose to it by more than the refinement tolerance. The informative split is the $26$/$48$ one---the gain appears where the blocker actually disturbs the aperture-to-user geometry and vanishes where it does not---which makes the design a no-regret substitute for focusing under blockage: it restores service where the edge geometry admits it and costs nothing elsewhere.}

\rev{The applicability boundary is physical, not statistical. An independent full-plane invariant confirms that a fully opaque plane spanning the aperture yields exactly zero received rate for \emph{every} transmitter-side scheme, whereas a half-space edge---even with blocked fraction approaching unity---still admits edge diffraction: Airy beamforming requires an accessible diffracting edge and does not restore propagation through a closed obstruction, and Fig.~\ref{fig:Conceptual} is drawn accordingly. Separately, the compact three-parameter family does not scale to many simultaneous users: under the declared service floor, only $2$ of $8$ three-user and $0$ of $4$ four-user multi-blocker stress rows are serviceable, so denser multiplexing requires enlarging the RF codebook or more flexible hardware.}

\section{Airy-Beam Parameter Optimization in Mixed Shadow-Bright Scenarios}
\label{sec:optimization}

While Section~IV demonstrated the efficacy of Airy beams in purely shadowed regions, a more complex challenge arises in mixed blockage scenarios, where some users are shadowed while others reside in the bright (LoS) region. In such cases, the extended oscillatory structure of the Airy beam---essential for its self-healing property---can cause severe interference to the LoS users. \rev{This section presents an interference-aware optimization of the three Airy parameters. The optimized beam does not steer an analog null onto the bright user---as the per-user field evidence below shows, it can even raise the raw cross-user leakage; its role is to improve the whitened weak mode and the ZF power cost of the multi-user channel, while the digital ZF stage performs the final interference cancellation.}

\subsection{Scenario: The Interference Challenge}

We consider a mixed geometry as depicted in \rev{Fig.~\ref{fig:Conceptual}}:
\begin{itemize}
    \item \textbf{UE-1 (Shadow):} Located at $(-5\lambda, 250\lambda)$, served by an Airy beam.
    \item \textbf{UE-2 (Bright - Hard Case):} Located at $(+3.5\lambda, 300\lambda)$. This position is termed a ``hard case'' because it lies close to the diffraction boundary, where the oscillating tail of UE-1's Airy beam is strong.
    \item \textbf{Precoding Strategy:} UE-2 uses a fixed traditional beam (optimal for LoS). The optimization focuses solely on UE-1's Airy beam parameters.
\end{itemize}

If we apply the Airy-Geo strategy (geometric steering) from Section~IV, the main lobe serves UE-1, but the side-lobes heavily pollute UE-2's channel. This degrades the \rev{whitened} conditioning of the channel, causing the ZF precoder to amplify noise and limiting the sum-rate. \rev{All quantitative results in this section are produced under the calibrated evaluation contract declared in Sec.~IV-D.}

\subsection{Problem Formulation: Joint Parameter Optimization}

To mitigate this coupling, we exploit the parameterized structure of the Airy beam. The complex aperture field depends on the tuple $\bm{\psi} = (B, F, \theta_1)$. Our goal is to find the optimal $\bm{\psi}^*$ that maximizes the sum-rate while maintaining a service guarantee for the shadowed user.

The optimization problem is formulated as:
\begin{subequations}
\begin{align}
    \max_{B, F, \theta_1} \quad & \Rsum(\Heff(B, F, \theta_1)) \\
    \text{s.t.} \quad & |h_{11}(\bm{\psi})|^2 \ge \eta \cdot |h_{11}^{\text{geo}}|^2,
\end{align}
\end{subequations}
where $h_{11}^{\text{geo}}$ is the channel gain obtained using geometric steering, and $\eta \in [0.3, 0.5]$ is a relaxation factor. This constraint ensures that the optimization does not trivially minimize interference by shutting down the target user's link.

\begin{algorithm}[!t]
\small
\caption{Coarse-to-Fine Airy \rev{Parameter} Optimization}
\label{alg:airy_null_opt}
\DontPrintSemicolon
\KwIn{Geometry; noise power $N_0$; transmit power $P_{\text{tx}}$; service constraint factor $\eta$; coarse grids $\mathcal{B}_c,\mathcal{F}_c,\Delta\Theta_c$; fine grids $\mathcal{B}_f,\mathcal{F}_f,\Delta\Theta_f$.}
\KwOut{Optimal UE-1 Airy parameters $(B^\star, F^\star, \theta_1^\star)$.}

\BlankLine
\textbf{Step 0: Initialization \& Baseline}\;
$\theta_1 \leftarrow \theta_{\text{geo}}$; compute baseline gain $|h_{11}^{\text{geo}}|^2$ and threshold $\tau \leftarrow \eta |h_{11}^{\text{geo}}|^2$\;
Fix UE-2 beam (traditional) and compute its channel column $\mathbf{h}_2$\;

\BlankLine
\textbf{Step 1: Coarse Search (Global)}\;
$R_{\text{best}} \leftarrow -\infty$\;
\ForEach{$B \in \mathcal{B}_c$}{
  \ForEach{$F \in \mathcal{F}_c$}{
    \ForEach{$\Delta\theta \in \Delta\Theta_c$}{
      Generate candidate UE-1 Airy beam $E_1(x)$ with parameters $(B, F, \theta_{\text{geo}}+\Delta\theta)$\;
      Propagate to obtain $\mathbf{h}_1$\;
      \If{$|h_{11}|^2 < \tau$}{\textbf{continue}\;}
      Form $\Heff \leftarrow [\mathbf{h}_1\ \mathbf{h}_2]$ and compute sum-rate $R$\;
      \If{$R > R_{\text{best}}$}{
        Update $(B_{\text{best}}, F_{\text{best}}, \theta_{\text{best}}) \leftarrow (B, F, \theta_{\text{geo}}+\Delta\theta)$\;
        $R_{\text{best}} \leftarrow R$\;
      }
    }
  }
}

\BlankLine
\textbf{Step 2: Fine Search (Local Refinement)}\;
Define fine grids around $(B_{\text{best}}, F_{\text{best}}, \theta_{\text{best}})$ to form $\mathcal{B}_f,\mathcal{F}_f,\Delta\Theta_f$\;
\ForEach{$B \in \mathcal{B}_f$}{
  \ForEach{$F \in \mathcal{F}_f$}{
    \ForEach{$\Delta\theta \in \Delta\Theta_f$}{
      Generate candidate UE-1 Airy beam $E_1(x)$ with parameters $(B, F, \theta_{\text{best}}+\Delta\theta)$\;
      Propagate to obtain $\mathbf{h}_1$\;
      \If{$|h_{11}|^2 < \tau$}{\textbf{continue}\;}
      Form $\Heff \leftarrow [\mathbf{h}_1\ \mathbf{h}_2]$ and compute sum-rate $R$\;
      \If{$R > R_{\text{best}}$}{
        Update $(B_{\text{best}}, F_{\text{best}}, \theta_{\text{best}}) \leftarrow (B, F, \theta_{\text{best}}+\Delta\theta)$\;
        $R_{\text{best}} \leftarrow R$\;
      }
    }
  }
}

\BlankLine
\textbf{Return:}\;
Set $(B^\star, F^\star, \theta_1^\star) \leftarrow (B_{\text{best}}, F_{\text{best}}, \theta_{\text{best}})$\;

\end{algorithm}

\subsection{Proposed Solution: Coarse-to-Fine Search}

Since the objective function is non-convex and oscillatory due to the phase structure of Airy beams, gradient-based methods may plunge into local optima. To address this, we propose a robust two-stage search strategy, summarized in Algorithm~\ref{alg:airy_null_opt}.

\subsection{Complexity Analysis}
The computational efficiency of the proposed beamforming strategy is a critical factor for practical implementation. Unlike fully digital precoding schemes that require optimizing complex-valued weights for every antenna element, the proposed coarse-to-fine Airy \rev{parameter} search optimizes only three scalar analog parameters $(B, F, \theta_1)$. 

Let $N_{\text{coarse}} = |\mathcal{B}_c||\mathcal{F}_c||\Delta\Theta_c|$ and $N_{\text{fine}} = |\mathcal{B}_f||\mathcal{F}_f||\Delta\Theta_f|$ denote the number of candidate parameter tuples evaluated in the coarse and fine search stages, respectively, consistent with the sets defined in Algorithm 1. For each candidate tuple, the dominant computational cost lies in the evaluation of the Fresnel propagation operator $\mathcal{P}_z$ (see Eq. (3)). Implementing this operator via the FFT-based Angular Spectrum Method on a transverse aperture grid of size $N_x$ (where $N_x \gg N$ is chosen to ensure sufficient sampling density and mitigate boundary reflections) yields a propagation complexity of $\mathcal{O}(N_x \log N_x)$. Consequently, the total computational complexity of the proposed algorithm scales as:
\begin{equation}
    C_{\text{total}} = \mathcal{O}\left( (N_{\text{coarse}} + N_{\text{fine}}) N_x \log N_x \right).
\end{equation}
It is worth noting that since the traditional focused beam for the Line-of-Sight (LoS) user (UE-2) is fixed during the optimization of the shadowed user (UE-1), its effective channel column $\mathbf{h}_2$ can be precomputed once. This reduces the per-iteration cost solely to the update of the Airy beam column $\mathbf{h}_1$.

\textit{Comparison with Conventional Schemes:} In contrast, a conventional fully digital multi-user beamforming optimization typically involves $N \times K$ complex variables. Solving for the optimal precoder often requires iterative algorithms (e.g., Weighted MMSE or SCA) involving matrix inversions, which scale with a complexity of at least $\mathcal{O}(K N^3)$ or $\mathcal{O}(K^2 N^2)$ per iteration. For ELAA systems where $N$ is large (e.g., $N \ge 256$), such operations are impractical for real-time. Our proposed method, leveraging the parameterized physics of Airy beams, drastically reduces the dimensionality of the search space. Furthermore, the optimization can be performed offline to generate a geometry-based codebook (Lookup Table), rendering the online complexity negligible ($\mathcal{O}(1)$) for static blockage scenarios.

\rev{\textit{Reference quality of the local refinement:} The coarse-to-fine search is benchmarked against a dense parameter grid. Over the $16$ serviceable reference scenes of the campaign, the $95$th-percentile signed gap between the geometry-initialized local search and the dense-grid reference is $0.053$~bit/s/Hz ($1.4\%$ in relative terms), with a median $37.6\times$ reduction in propagation evaluations; the local search even beats the finite grid in $2$ of $20$ rows, so the grid serves as a reference rather than a strict upper bound. Four fully blocked scenes are total outages for every scheme; they belong to the applicability boundary of Sec.~IV-D rather than to the refinement statistics.}

\begin{figure*}[!t]
    \centering
    \includegraphics[width=.92\textwidth]{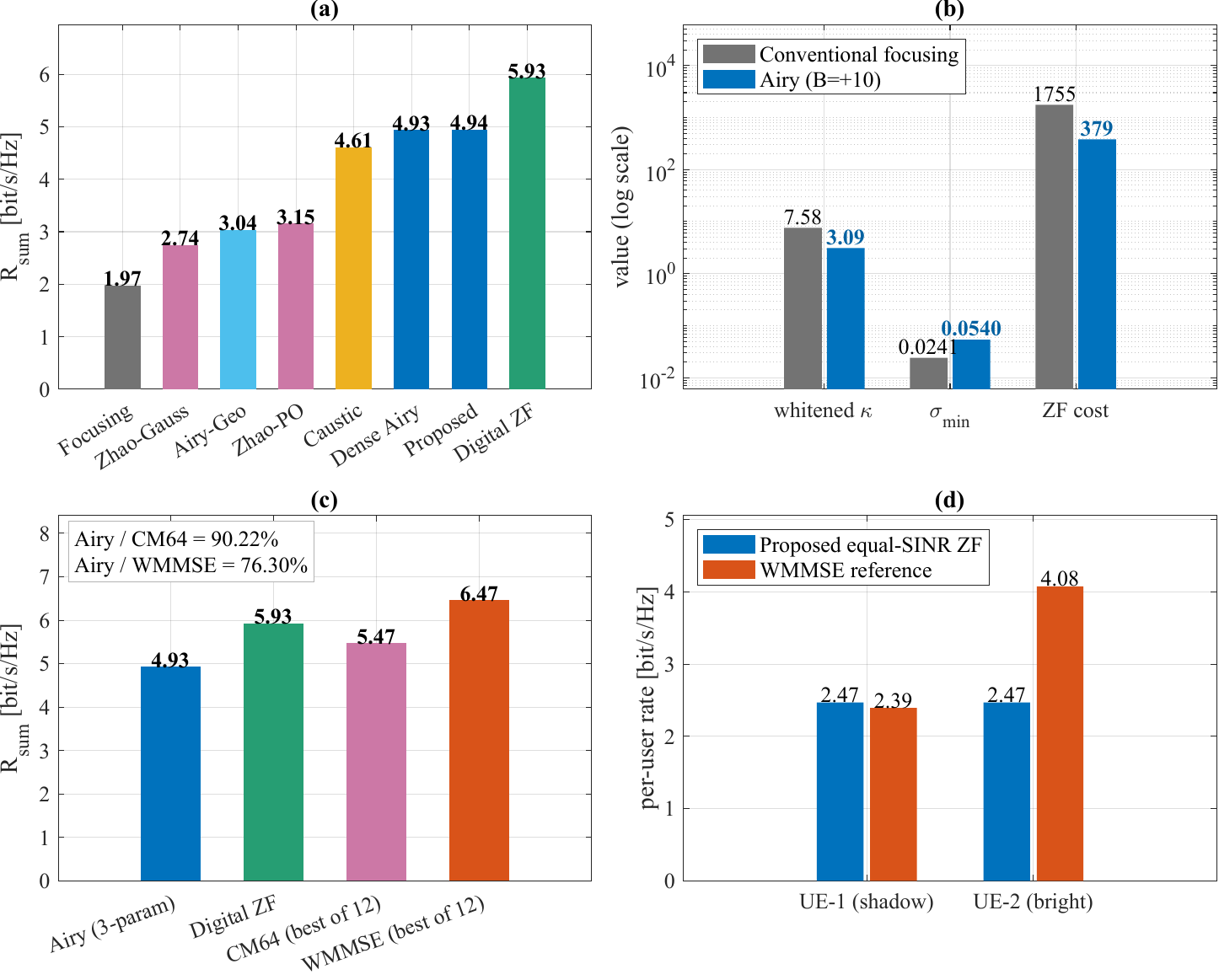}
    \caption{\rev{Mixed-scene evidence under one channel, noise, and total-power contract. (a)~Structured and digital references. (b)~Gram-whitened separability from focusing to the optimized Airy member: $\kappa$ falls $7.58\to 3.09$, $\sigma_{\min}$ rises $0.0241\to 0.0540$ ($2.24\times$), and the ZF power cost $\sum_i \sigma_i^{-2}$ falls $1755\to 379$ ($-78.4\%$). (c)~Achieved optimization references; the WMMSE and CM64 values are the best achieved over $12$ declared starts. (d)~Per-user rates: WMMSE attains its aggregate advantage without starving the shadowed user, while the proposed equal-SINR ZF design gives the minimum-rate user slightly more.}}
    \label{fig:canonical_results}
\end{figure*}

\subsection{Mixed-Scene Results: Service Rebalancing Through Conditioning, Not Analog Nulls}

\rev{\textbf{Optimized three-parameter design.} Under the total-power constraint of Sec.~II-C, the reproducible optimum of the three-parameter Airy family in the mixed scene is $(B,F,\Delta\theta)=(+10,\,1.9~\mathrm{m},\,+1.9^{\circ})$; all numbers and figures in this section are produced under this constraint.}

\rev{Fig.~\ref{fig:canonical_results}(a) compares the schemes in the mixed scene under the common contract. Conventional focusing yields $1.97$~bit/s/Hz. The literature-inspired constructions---the closed-form Gaussian and phase-only Airy mappings following \cite{Zhao2025DataCenter} and the caustic-trajectory design following \cite{Guerboukha2024Curving}, each source-audited and re-implemented under the identical aperture, channel, noise, and total-power convention---reach $2.74$, $3.15$, and $4.61$~bit/s/Hz, respectively. The proposed geometry-initialized and locally refined design attains $4.94$~bit/s/Hz, and a dense-grid member of the same family attains $4.93$~bit/s/Hz, confirming that the reported gain is not a grid-density artifact. Unrestricted digital ZF reaches $5.93$~bit/s/Hz. The physics-informed learning framework of \cite{ref10} addresses the same blockage problem, but no executable trained model is distributed with it, so a like-for-like comparison is not reproducible and is not attempted.}

\rev{Fig.~\ref{fig:canonical_results}(b) reports the mechanism with the whitened metrics of Sec.~II-D: from focusing to the optimized Airy member, $\kappa(\bar{\mathbf{H}})$ falls from $7.58$ to $3.09$, the weak-mode strength $\sigma_{\min}(\bar{\mathbf{H}})$ rises from $0.0241$ to $0.0540$ (a factor of $2.24$), and the ZF power cost drops from $1755$ to $379$ ($-78.4\%$, i.e., $4.6\times$ less noise enhancement at the same radiated power).}

\begin{figure*}[!t]
    \centering
    \includegraphics[width=.92\textwidth]{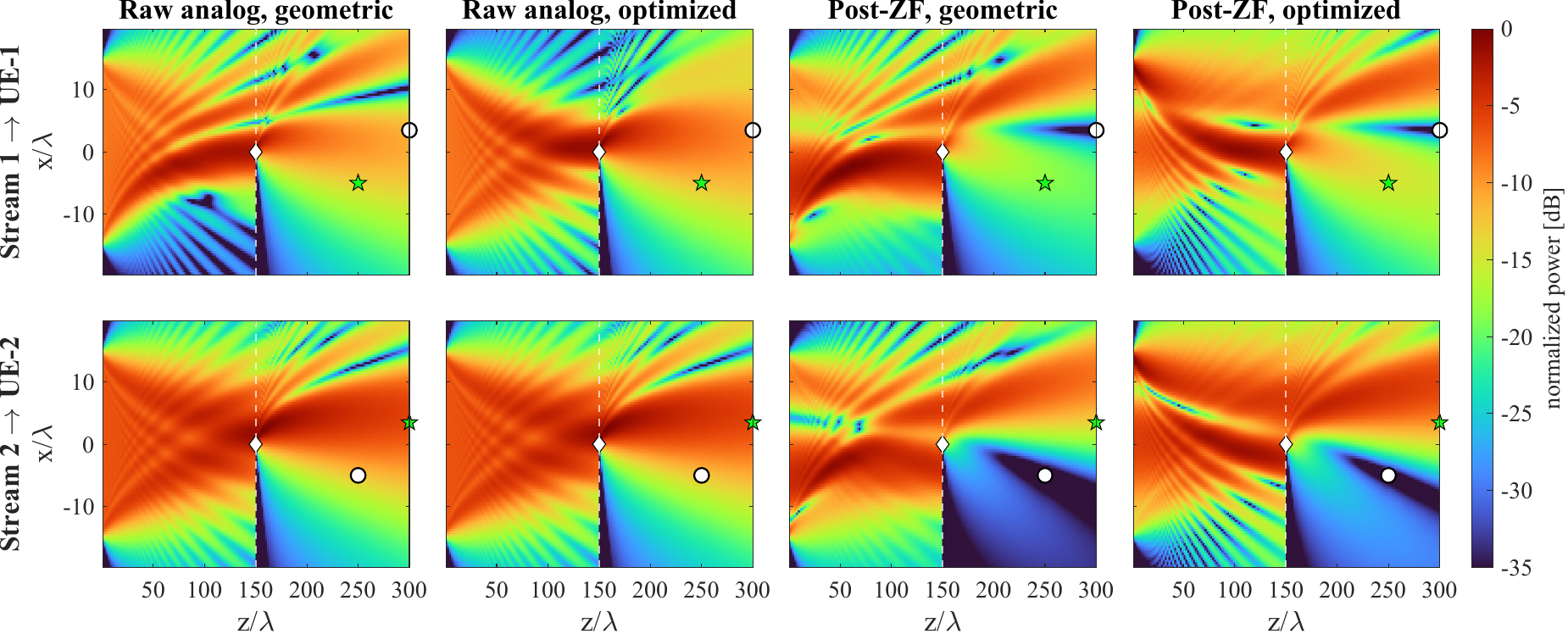}
    \caption{\rev{Per-user field evidence for the mixed scene. Columns: raw analog fields of the geometric and optimized designs, then post-ZF stream fields of the two designs; rows: stream~1 (intended for the shadowed UE-1, star) and stream~2 (intended for the bright UE-2); each raw/post-ZF pair is normalized to its own peak (color scale $-35$ to $0$~dB; the circle marks the cross user, the diamond the obstacle edge). From the geometric to the optimized Airy design, the raw desired endpoint power rises $4.94\to 5.27$ while the raw cross-user leakage \emph{also} rises $2.51\to 3.39$; after digital ZF the residual leakage is numerically zero. The RF stage is thus not an analog null former; it is a conditioning and shadow-energy stage, and the digital stage performs the final cancellation.}}
    \label{fig:blocked_scene_users}
\end{figure*}

\rev{Fig.~\ref{fig:blocked_scene_users} separates the per-user endpoint fields into raw analog and post-ZF stream components. Going from the geometric to the optimized Airy beam, the raw desired endpoint power rises from $4.94$ to $5.27$, but the raw cross-user leakage \emph{also rises}, from $2.51$ to $3.39$: the optimized beam does not place an analog null on the bright user. After the digital stage, the residual leakage is numerically zero and the desired post-ZF endpoint power more than doubles. The mechanism is therefore \emph{service rebalancing through conditioning}: the RF stage strengthens the whitened weak mode that blockage had collapsed, and the digital ZF stage spends the recovered margin on interference cancellation.}

\begin{figure*}[!t]
    \centering
    \includegraphics[width=.98\textwidth]{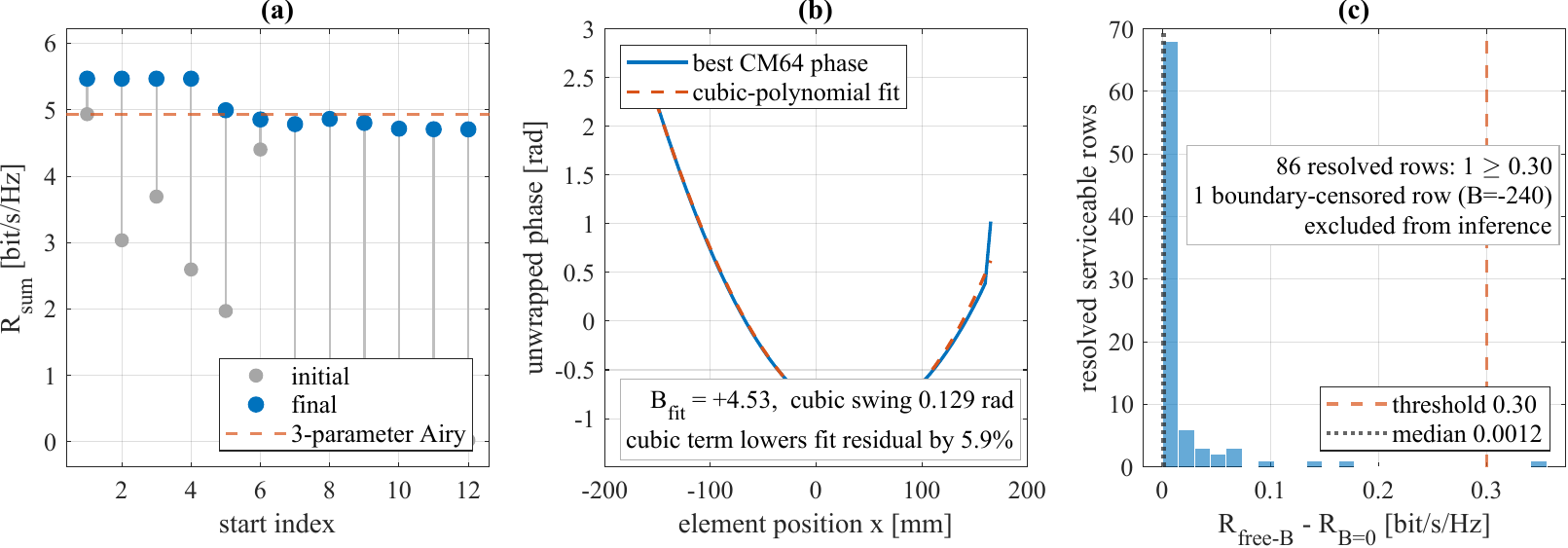}
    \caption{\rev{Optimization-reference evidence. (a)~The constant-modulus $64$-phase (CM64) reference from $12$ deterministic starts: best value $5.47$~bit/s/Hz against $4.93$ for the three-parameter family ($90.2\%$). (b)~The best CM64 phase profile against a cubic-polynomial fit: the fitted curvature is $B_{\mathrm{fit}}=+4.53$ with a cubic swing of only $0.129$~rad ($5.9\%$ residual reduction), so the free optimum is \emph{not} significantly Airy-shaped. (c)~Within-family cubic ablation across the breadth campaign: the median resolved gain of free~$B$ over $B=0$ is $0.0012$~bit/s/Hz, and one of $86$ resolved serviceable rows reaches the pre-declared $0.30$ threshold; one boundary-censored row ($B=-240$) is disclosed and excluded from inference.}}
    \label{fig:optimization_limits}
\end{figure*}

\subsection{Distance From Unconstrained Optimization}

\rev{Two references quantify how much performance the three-parameter restriction gives up; both are the best values achieved by declared multi-start procedures.}

\rev{\emph{Fully digital relaxation (WMMSE).} Every hybrid product $\mathbf{F}=\mathbf{W}_{\text{RF}}\mathbf{W}_{\text{BB}}$ is feasible for an unconstrained fully digital precoder under the same power expression, because $\|\mathbf{W}_{\text{RF}}\mathbf{W}_{\text{BB}}\|_F^2=\operatorname{tr}(\mathbf{W}_{\text{BB}}^H\mathbf{G}\mathbf{W}_{\text{BB}})$; the optimum of the relaxed fully digital problem therefore upper-bounds every hybrid design considered here. Solving the relaxed problem with $12$-start iterative WMMSE achieves $6.47$~bit/s/Hz, so the proposed design reaches $76.3\%$ of this optimization-based fully digital upper reference (Fig.~\ref{fig:canonical_results}(c)). The comparison is also fair to the shadowed user (Fig.~\ref{fig:canonical_results}(d)): WMMSE serves both users ($2.39$ and $4.08$~bit/s/Hz), and the proposed equal-SINR ZF design gives the minimum-rate user slightly more ($2.47$~bit/s/Hz); the WMMSE aggregate advantage comes from the bright user, not from abandoning the weak one.}

\rev{\emph{Constant-modulus $64$-phase reference (CM64).} Removing the three-parameter restriction while keeping the constant-modulus aperture, the fixed bright-user beam, and the same channel and evaluator, a $64$-variable phase optimization from $12$ deterministic starts (with the Airy solution included as one start, so this class strictly contains the proposed family) achieves $5.47$~bit/s/Hz (Fig.~\ref{fig:optimization_limits}(a)). Three physical parameters thus recover $90.2\%$ of the best achieved element-wise design. To answer the natural follow-up question---does the unconstrained optimum look like an Airy beam?---Fig.~\ref{fig:optimization_limits}(b) fits a cubic polynomial to the best CM64 phase profile: the cubic contribution amounts to a $0.129$-rad swing that lowers the fit residual by only $5.9\%$. The free optimum is \emph{not} significantly Airy-shaped; the value of the Airy parameterization is dimensional efficiency rather than shape optimality.}

\rev{\emph{What the cubic degree of freedom is worth.} Within the family itself, an ablation of the cubic term (free~$B$ versus $B=0$ under identical search treatment) yields a principal-scene increment of $0.148$~bit/s/Hz, about $5\%$ of the total improvement over focusing. Across the breadth campaign, the median resolved increment is $0.0012$~bit/s/Hz and exactly one of $86$ resolved serviceable rows reaches the pre-declared $0.30$-bit/s/Hz threshold (Fig.~\ref{fig:optimization_limits}(c)); one row that terminated on an expanded search rail at $B=-240$ is disclosed and excluded from the distributional inference. Most of the blockage gain is produced by re-steering and re-focusing; the cubic term is a small, regime-dependent extra coordinate.}

\subsection{Operating-Boundary Diagnostics}

\rev{\emph{Practical distance range.} Sixteen pre-specified range anchors span $z_1/\lambda=175$ to $2400$, i.e., $1.87$--$25.7$~m or $0.092$--$1.26$ Fraunhofer distances. Under fixed transmit power, the anchors split into $10$ advantage, $5$ equivalent, and $1$ failure outcomes, and a calibrated eligibility map admits $29$ of $57$ evaluated rows under joint geometry and relative-service criteria. Because the optimizer branches are non-monotonic in range, these samples define a tested operating region with explicit boundaries rather than a universal maximum range.}

\rev{\emph{Obstacle scattering.} The opaque-screen model includes forward propagation and edge diffraction but excludes material reflection, finite-thickness penetration, and diffuse scattering. Adding an identical scattered-channel perturbation to every scheme at relative levels $\eta=-20$, $-10$, and $0$~dB produces searched proposed-rate ranges of $4.24$--$5.54$, $0.54$--$7.39$, and $6.4\times10^{-5}$--$9.86$~bit/s/Hz, respectively. These finite deterministic samples are illustrative rather than worst-case bounds: they show that unmodeled multipath can help or harm the link.}

\rev{\emph{Wideband behavior.} With the RF phases designed once at $f_c=28$~GHz ($\lambda\approx 10.7$~mm) and the propagation, Gram matrix, and digital stage recomputed on $33$ subcarriers under fixed total in-band signal and noise power, no band-mean rate-loss crossing occurs up to the tested $20\%$ fractional bandwidth; the supported bandwidth is therefore reported as right-censored ($\geq 20\%$ tested), not as a maximum, consistent with the bandwidth considerations highlighted in \cite{SinghSPM}. At $20\%$, the mixed scene retains a worst-user band-mean rate of $2.46$~bit/s/Hz (worst tone $2.12$), while the double-shadow scene remains poor in absolute terms ($0.055$ band mean). Separately, the $d=0.49\lambda_c$ element spacing implies a conservative alias-free spatial-sampling ceiling of $4.1\%$ fractional bandwidth; results beyond that ceiling are dispersion diagnostics rather than alias-free hardware operation.}

\rev{\emph{Relation to RIS.} Airy beamforming and RIS address different path mechanisms and are complementary rather than interchangeable. The full-plane invariant of Sec.~IV-D fixes the boundary: with no accessible edge, no transmitter-side scheme survives, whereas a deployed reflecting surface can create a separate reflected route. Transmitter-side edge diffraction requires no deployed surface; RIS assistance covers geometries that offer no usable edge, at the cost of surface hardware, channel acquisition, control, and latency overhead. A systematic comparison of the two mechanisms over surface placement, aperture, material loss, phase quantization, control delay, multipath, and bandwidth is left for future work.}

\section{Conclusion and Future Work}
\label{sec:conclusion}

In this paper, we have investigated the fundamental limitations of traditional radiative near-field (RNF) focusing in half-space blockage scenarios and proposed a novel Airy-beam-assisted transmission strategy to overcome these challenges. Our analysis, grounded in rigorous angular-spectrum diffraction modeling, revealed that conventional focused beams suffer from a catastrophic ``singular value collapse'' when encountering knife-edge obstacles, rendering zero-forcing precoding ineffective for shadowed users.

\rev{To address this, we exploited the auto-bending property of Airy beams within a two-stage architecture: a three-parameter analog family with geometry-based initialization and local refinement, followed by digital ZF under one declared total-radiated-power contract. In the principal mixed scene the design attains $4.94$~bit/s/Hz against $1.97$ for conventional focusing, raises the whitened weak singular mode by a factor of $2.24$, and cuts the ZF power cost by $78\%$; across $36$ blocked-user geometries, whitened conditioning and minimum user rate improve in every scene. Benchmarked against stronger references, three physical parameters recover $90.2\%$ of the best achieved $64$-phase constant-modulus design and $76.3\%$ of an iterative fully digital WMMSE reference, while giving the shadowed user a slightly higher rate than WMMSE. The boundaries are stated with equal weight: an accessible diffracting edge is required (a fully opaque plane defeats every transmitter-side scheme), only $2$ of $8$ three-user and $0$ of $4$ four-user stress scenes meet the declared service floor, and the cubic coordinate contributes a small, regime-dependent share of the gain.}

\rev{This work positions Airy beamforming as a hardware-efficient, transmitter-side option complementary to RIS.} Future research directions include extending this framework to 3D blockage geometries, investigating the integration of Airy beams with dynamic user tracking, and exploring learning-based methods for real-time beam parameter adaptation in rapidly changing environments.

\vfill

\end{document}